\documentclass[aps,twocolumn,pra,tightenlines,floatfix,showpacs,superscriptaddress]{revtex4-1}
\usepackage{graphicx}
\usepackage{amsmath}
\usepackage{amssymb}
\usepackage{times}
\usepackage{epstopdf}
\usepackage[english]{babel}
\usepackage{hyperref}

\newcommand{\bracket}[2]{\ensuremath{\left\langle #1 \middle| #2 \right\rangle}}

\begin{document}

\title{Protocols for dynamically probing topological edge states and dimerization with fermionic atoms in optical potentials}

\author{Mekena Metcalf}
\affiliation{School of Natural Sciences, University of California, Merced, CA 95343, USA}
\author{Chen-Yen Lai}
\affiliation{School of Natural Sciences, University of California, Merced, CA 95343, USA}
\author{Kevin Wright}
\affiliation{Dept. of Physics and Astronomy, Dartmouth College, Hanover NH 03755, USA}
\author{Chih-Chun Chien}
\affiliation{School of Natural Sciences, University of California, Merced, CA 95343, USA}

\begin{abstract}
Topological behavior has been observed in quantum systems including ultracold atoms. However, background harmonic traps for cold-atoms hinder direct detection of topological edge states arising at the boundary because the distortion fuses the edge states into the bulk. We propose experimentally feasible protocols to probe localized edge states and dimerization of ultracold fermions. By confining cold-atoms in a ring lattice and changing the boundary condition from periodic to open using an off-resonant laser sheet to cut open the ring, topological edge states can be generated. A lattice in a topological configuration can trap a single particle released at the edge as the system evolves in time. Alternatively, depleting an initially filled lattice away from the boundary reveals the occupied edge states. Signatures of dimerization in the presence of contact interactions can be found in selected correlations as the system boundary suddenly changes from periodic to open and exhibit memory effects of the initial state distinguishing different configurations.
\end{abstract}


\maketitle

Emergent condensed matter phenomena arising from topological behavior has incited increasing attention. For instance, topological insulators can host edge states within an insulating energy gap and are characterized by topological invariants arising from their band structures~\cite{Asboth2015, ShenTI, Chiu2016, Zhang-TIRev, Kane_TIRev}. The versatility of ultracold atoms for simulating many-body systems~\cite{Chien15, Bloch-QuantSimRev} leads to realizations of important topological systems such as the Su-Schrieffer-Heeger (SSH) model~\cite{Atala2013}, Harper-Hofstadter Hamiltonian~\cite{HarperHam, HofHam}, and the elusive Haldane model~\cite{HaldaneHam}.
Topological structures also support unique dynamical features realizable with cold atoms such as quantized charge transport~\cite{Nakajima2016, Lohse2016}.

\begin{figure}
\includegraphics[width=\columnwidth]{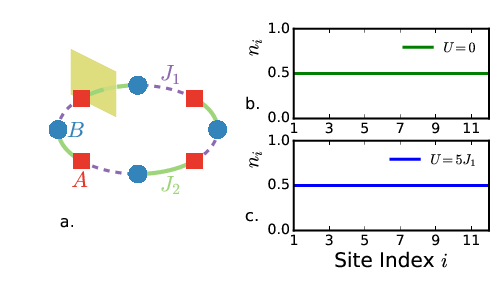}
\caption{(a) Trapping set up with optical ring lattice cut with an off resonant laser sheet (yellow) to open boundaries. The lattice has unit cell with two sites A (red square) and B (blue circle). Density profile for a half-filled lattice with $L\!=\!12$ sites, half filling, $J_1\!/\!J_2\!=\!0.5$, and interaction strength (b) $U\!/\!J_1\!=\!0$ and (c) $U\!/\!J_1\!=\!5$.}
\label{Fig:RingLattice}
\end{figure}

The SSH model has been constructed with an optical superlattice in a confining potential to probe quantum phases~\cite{Atala2013}. It supports nontrivial band topology with sublattice symmetry and localized edge states when the system is uniform. Direct detection of edge states, however, has been hindered by the confining potential, which distorts the density profile and prevents a sharp boundary.There have been proposals for observing topological edge modes in 2D cold-atom systems~\cite{Goldman13}. Here, an alternate setup in 1D with well defined boundaries, where edge states can be generated and detected, is proposed. Rather than confining atoms in a harmonic trap, we propose that they be confined in an dimerized ring lattice formed by a patterned optical dipole trap. If the tunneling between a selected pair of sites is turned off, the ring is effectively cut open, resulting in a one dimensional lattice with open boundary conditions as illustrated in Figure~\ref{Fig:RingLattice}(a). This system could be realized with some adaptation of current ``Fermi gas microscope'' experiments~\cite{MullerPhD2011,Greiner2015,Cheuk2015}, which we will describe below.  


Cutting open the ring lattice will generate localized edge states in real space, but additional methods are needed to distinguish these occupied edge states from the background of particles in other states. For example, the density profile of a half-filled, open lattice is featureless both for noninteracting and interacting systems, as shown in Figures~\ref{Fig:RingLattice}(b) and~\ref{Fig:RingLattice}(c). We propose to detect the edge states using two specific dynamical protocols: an injection method and a depletion method. A single fermion injected at the edge of an empty open lattice with nontrivial topology does not fully spread out as time evolves, because most of its weight is kept by the localized state at the edge. A recent experiment used a similar idea, injecting bosons onto the boundary of a lattice constructed in momentum space~\cite{Meier2016}. Our proposal is a real-space counterpart of this experiment. Another way to probe the edge states is to dynamically remove atoms~\cite{Chern2014}. Removing atoms from the region near the edge over time depletes most of the states except the edge states localized at the boundary, allowing them to be distinguished from the bulk states. 

Noninteracting topological systems have been quite thoroughly characterized, but identifying and characterizing topological properties in interacting systems remains a great challenge~\cite{Chiu2016}. Topological invariants, calculated from the single-particle wavefunction, are more difficult to evaluate for many-body states. While some invariants can be calculated by Green's functions in the thermodynamic limit~\cite{Manmana2012, VolovikBook}, cold-atom systems are usually of finite size and require more direct characterization. A half-filled SSH model with an even number of sites has particle-hole symmetry, and has been shown to remain topological in the presence of interactions~\cite{Hatsugai2006, Guo2011}. A global topological number may also be determined by adding a phase twist to the boundary and analyzing the phase change when interactions are present~\cite{Guo2011, Hirano2008, Guo2015}. However, the structure of many-body topological edge states remains elusive.


Instead of mapping out the edge state in interacting systems, we propose analyzing selected correlation functions for revealing the dimerized structures in the SSH model with interactions. We found that that finite correlations will survive across a broken link in a ring lattice after a boundary is created by inserting a laser sheet. The evolution of spin-spin correlations can thus distinguish between trivial and nontrivial topological configurations with and without interactions.

{\it Theoretical Background:} The SSH model describes a one-dimensional (1D) dimerized lattice with alternating hopping strengths~\cite{Asboth2015, ShenTI}. The 1D chain hosts two sites (A and B) in each unit cell. The Hamiltonian of an $N$-site noninteracting system is
\begin{equation}\label{eq:HSSH}
H = -\sum_{i=0}^{N}\Big[ J_1c_{B,i}^{\dagger}c_{A,i} + J_2c_{A,i+1}^{\dagger}c_{B,i} + h.c.\Big].
\end{equation}
Here $c_i^{\dagger }(c_i)$ represents the creation (annihilation) operator on site $i$. The hopping coefficients alternate with $J_1$ and $J_2$ as shown in Figure~\ref{Fig:RingLattice}(a). Depending on the boundary condition, the hopping between site $N$ and site $1$ is present (for periodic boundary condition) or absent (for open boundary condition). The unit of time $t_0\!=\!\hbar/J_1$ is set by the intra-cell hopping coefficient and $\hbar\!=\!1$. We first consider a periodic system and by taking a Fourier transform of the creation (annihilation) operators, $c_{i,A}\!=\!\sum_{k} a_k e^{ik\cdot x_i}$ and $c_{i,B}\!=\!\sum_{k} b_k e^{ik\cdot (x_i+a/2)}$, the Hamiltonian in momentum space is
$H(k)\!=\!
\sum_{k} \begin{pmatrix}
a_k^{\dagger} & b_k^{\dagger}
\end{pmatrix}
h(k)
\begin{pmatrix}
a_k\\
b_k
\end{pmatrix}
$ with
$h(k) =  \left( \begin{array}{cc}
0 & J_1e^{ika/2} + J_2e^{-ika/2}  \\
J_1e^{-ika/2}+J_2e^{ika/2} & 0
\end{array} \right).$
The lattice constant $a$ serves as the length unit and the momentum summation is over the first Brillouin zone.
The matrix $h(k)$ can be diagonalized yielding 
the energy dispersion $\epsilon(k)\!=\!\pm \sqrt{J_1^2+J_2^2+2J_1J_2\cos(k)}$.
A gap opens between the upper and lower bands in the bulk when $J_1\!\neq\!J_2$.

Topological insulators can be classified according to their symmetries~\cite{Chiu2016, ShenTI, Zhang-TIRev}.
The SSH model has chiral (sublattice) symmetry with an associated topological invariant called the winding number $\nu$ defined for systems with periodic boundary condition.
$\nu$ takes values of $\pm 1$ depending on whether $J_1/J_2\!>\!0$ or $J_1/J_2\!<\!0$, revealing the parameters of the Hamiltonian wrap around a point counterclockwise (clockwise) in the parameter space.
For systems with open boundary condition, zero-energy edge states localized at the boundary can emerge depending on the configuration.
The bulk-boundary correspondence~\cite{Asboth2015, Zhang-TIRev, Kane_TIRev} states that the number of edge modes is determined by the bulk topological invariant.

Here we propose measuring consequences of band topology by probing the topological edge states. An open boundary introduces an interface between the topological SSH chain and the topologically trivial vacuum, so edge modes can emerge if the configuration is suitable. Location of the edge states on sublattice A (sublattice B) depends on the condition $J_1/J_2\!<\!1$ ($J_1/J_2\!>\!1$) and sign of the associated winding number. For instance, in a long chain with one end ending with a $J_1$ link and $J_1/J_2 < 1$, the probability amplitude of an edge state $|\phi\rangle$ has the form
\begin{equation}\label{eq:edgemode}
\bracket{i}{\phi} = \mathcal{N}
\begin{pmatrix}
1, 0, -\frac{J_1}{J_2}, 0, \Big(-\frac{J_1}{J_2}\Big)^2,0,\Big(-\frac{J_1}{J_2}\Big)^3,...\\
\end{pmatrix}
\end{equation}
and it localizes on sublattice A at the boundary.
$\mathcal{N}\!=\![1- (\frac{J_1}{J_2})^2]^{1/2}$ is a normalization factor.
Similarly, an end with a $J_2$ link with $J_1/J_2\!>\!1$ can support an edge state with similar structures on the B sublattice (but with $J_2/J_1$ in the power series).
Therefore, for a chain with odd number of sites, one edge state is guaranteed and its location depends on the ratio of $J_1/J_2$. For a chain with even number of sites, there can be zero or two edge states. For instance, an open chain ending with $J_1$ links on both sides and $J_1/J_2\!<\!1$ supports an edge state on both sublattices, and the two modes hybridize to give one symmetric and one anti-symmetric edge mode. On the contrary, the same chain with $J_1/J_2\!>\!1$ yields no zero-energy edge states. For a ring lattice, an even number of sites is a natural setting before a laser sheet cuts it open, so we will refer to the case with two edge states after it is cut open as the topologically nontrivial and the one with no edge state as the topologically trivial. It is possible, however, to generate a ring with an odd number of lattices~\cite{Lee2014, KumarDopplerRing}.

In cold-atom experiments, usually accompanying the alternating hopping in superlattices is the addition of alternating onsite energies, which adds a term $\sum_{i}(-1)^i \Delta c_i^{\dagger}c_i$ to the Hamiltonian~\cite{Atala2013}.
The edge-state configuration is still maintained in the presence of alternating onsite energies as one can verify directly, although the edge-state energy is shifted by $\Delta$. For quantized charge pumping of particles~\cite{ShenTI, Nakajima2016, Lohse2016}, this addition of alternating onsite energies is essential because the edge states will fuse into the bulk~\cite{Bermudez2009} as the crossing $J_1\!=\!J_2$ happens when they are tuned dynamically. Furthermore, $\Delta$ can be tuned to zero~\cite{Atala2013}, so for simplicity we do not include the alternating onsite energy in our discussion.

{\it Experimental Challenges and Proposed Solutions:}
Alternating onsite energies do not destroy the edge states, but a confining harmonic trap can be detrimental to them.
A harmonic confining trap, $V(r)\!\propto\!r^{\gamma}$ with $\gamma\!=\!2$, keeps atoms in the optical setup, but it has been shown that if $\gamma\!<\!10$, the energy gap of a dimerized lattice closes due to distortion of the density profile and the edge states fuse into the bulk of the lattice~\cite{Atala2013}. When $\gamma$ is large the potential takes the form of a box potential with well defined boundary. Three dimensional box potentials created by intersecting a hollow tube beam with two sheet beams provide sharp boundaries~\cite{Gaunt2013}, but adding lattice potential on top of the box potential can be more challenging.
\begin{figure}
\includegraphics[width=\columnwidth]{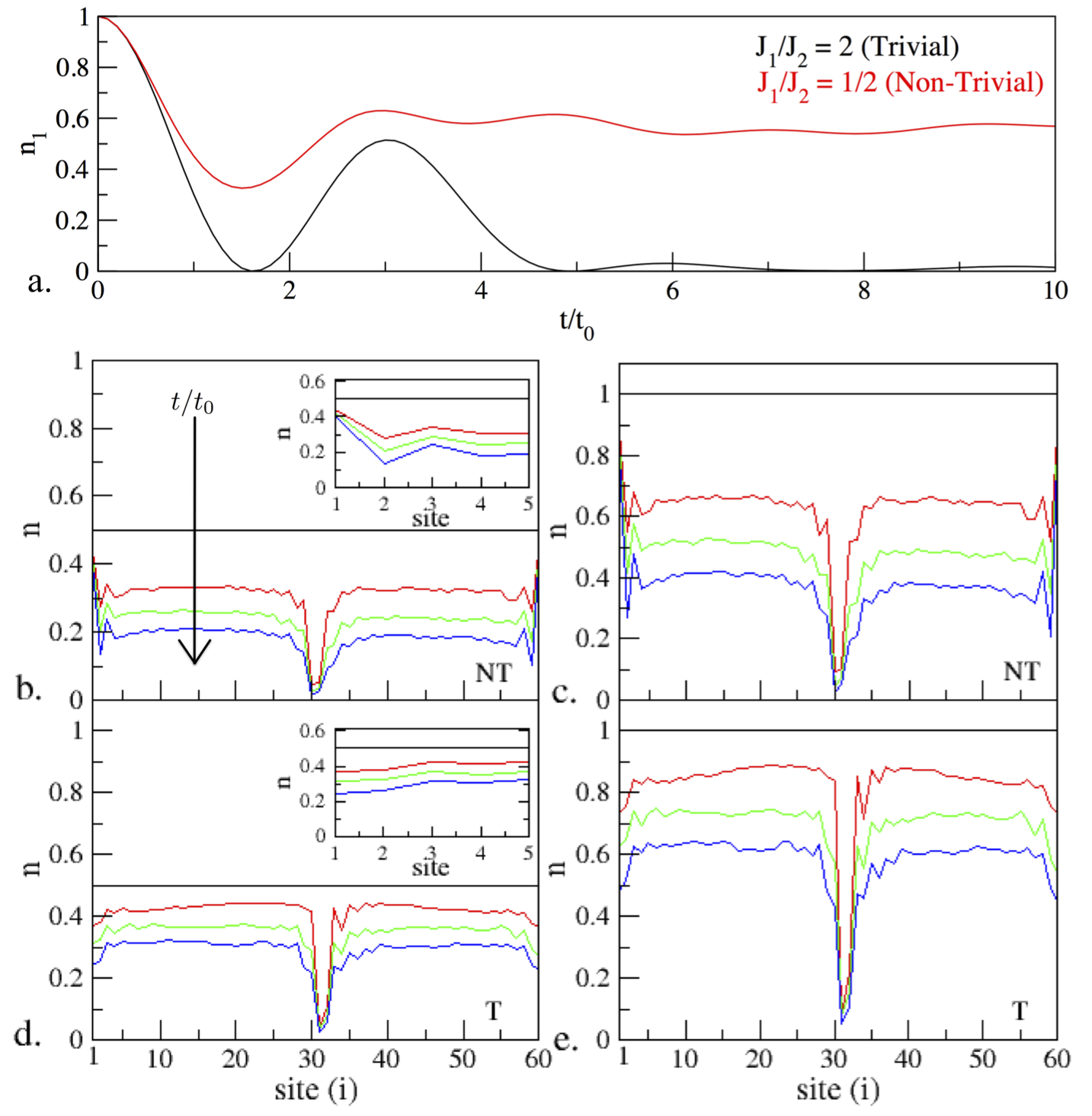}
\caption{(a) Density of the left-most site for a 60-site lattice with one fermion initially place on left boundary.
The topologically nontrivial case (red, $J_1/J_2\!=\!0.5$) has an edge mode retaining the injected particle while the injected fermion propagates away in the topologically trivial case (black, $J_1/J_2\!=\!2$).
(b) - (e) Evolution of the density profiles of a $60$-site lattice with a depletion beam removing particles near the center of the system ($i\!=\!31$).
The density profiles at selected times (from top to bottom) $t/t_0\!=\!0, \!50, \!100, \!150$ are shown.
The elements $\langle c_{31}^{\dagger}c_j \rangle$ and $\langle c_j^{\dagger}c_{31} \rangle $ for all $j$ are wiped out every $t_0/2$.
(b) and (c): The topologically nontrivial configuration ($J_1/J_2\!=\!0.5$, labeled "NT") displays prominent localized edge-state structures with higher density at the two open ends as the overall density is depleted.
(d) and (e): No localized structures emerge in the topologically trivial configuration ($J_1/J_2\!=\!2$, labeled "T"). The insets of (b) and (d) show the density profiles of the first five sites at the boundary.}
\label{Fig:Depletion}
\end{figure}

We propose a method for confining the atoms in a ring lattice cut with a laser sheet to open the boundary Figure~\ref{Fig:RingLattice}(a).
The appearance of edge states depends on where the link is cut.
If $J_1/J_2\!<\!1$ and the laser sheet cuts a strong link, $J_2$, two edge states emerge along with further correlation effects.
The laser sheet may introduce an energy shift on nearby sites. Assuming a shift of $\epsilon$ on the boundary site, the edge mode wavefunction \eqref{eq:edgemode} starts acquiring values $\epsilon/J_1, \epsilon/J_1 (-J_2/J_1), \cdots$ on those sites originally with zero amplitude. The edge mode survives in a chain with length less than $(1/2)[\ln(\epsilon /J_1)/\ln(J_1/J_2)]$. Making the laser sheet sharper to reduce the shift of onsite energy near the boundary will help the edge mode be more prominent. Interestingly, when $\epsilon = J_1$, the edge mode is completely destroyed, a result consistent with a direct mathematical evaluation~\cite{Kouachi06}. We will assume the collateral energy shift is negligible.

{\it Dynamical Signatures in Topological Noninteracting Systems:}
While edge-state contribution can be measured from transport in electronic systems~\cite{Kane_TIRev,Zhang-TIRev}, there is a need to develop methods to measure the edge states in cold-atom systems~\cite{Chien15}.
Bosonic atoms revealed characteristic edge states in momentum space using a single site injection method~\cite{Meier2016}.
Here we found that real-space single site injection also serves as a suitable method to probe edge states.
The idea is to have an empty lattice and inject a single atom onto the site at the open boundary, then the particle propagates in time.
The time-evolved density profile can be monitored by evolving the correlation matrix $\langle c_i^{\dagger}c_j \rangle$ according to the equation of motion
\begin{equation}
-i\frac{d\langle c_i^{\dagger}c_j \rangle}{dt} = \langle[ c_i^{\dagger}c_j, H]\rangle.
\end{equation}
The initial condition is $\langle c_1^{\dagger}c_1\rangle (t\!=\!0)\!=\!1$ and all other elements are zero, where the first site is at the open boundary.
The density on site $i$ is $\langle c_i^{\dagger}c_i \rangle$ and can be extracted from the time-evolved correlation matrix.
A fourth-order Runge-Kutta algorithm is used to numerically evaluate the time-evolution.

A single particle injected onto an end site not supporting an edge state propagates away, while a significant fraction of the wavefunction will be retained if there is an edge mode at the boundary.
This can be understood by Eq.~\eqref{eq:edgemode} because the edge mode has weight $\mathcal{N}^2=1-(J_1/J_2)^2$ at the boundary site, so the injected fermion will fill the edge mode significantly.
Figure~\ref{Fig:Depletion}(a) contrasts the remaining density on the injection site with and without an edge mode.
For the topologically nontrivial case, the edge mode retains $\approx 0.6$ of the weight, which is below the $\mathcal{N}^2=0.75$ upper bound.
The discrepancy is mostly due to finite-size effects as well as the small local spreading of the edge mode.
In contrast, the density practically vanishes as time evolves in the topologically trivial case.
The remaining density thus distinguishes the trivial and non-trivial topological configurations, providing measurable evidence of edge states in real space.

The method of depletion provides another measurable evidence of topological edge states, and a similar depletion method has been considered for an optical kagome lattice~\cite{Chern2014}.
By depleting the atoms away from the boundary over time, density of localized particles on the edge become prominent because their weights are small in the bulk.
To deplete atoms, a focused laser can eject atoms from the optical trap~\cite{Caliga2016}.
Alternatively, an electronic beam has been used to deplete bosons since, excited through collisions, atoms can escape the trap~\cite{Barontini2013}.
We consider a uniformly filled lattice with open boundary condition as the initial state and approximate the depletion of atoms on site $m$ by removing all correlations $\langle c_m^{\dagger}c_i \rangle = \langle c_i^{\dagger}c_m \rangle = 0$ associated with the depletion site.
Interestingly, depleting with a finite removal rate is necessary, or else the depleting beam will act as a barrier preventing particle transport.
Here we use a removal rate of depleting the selected correlations every $t_0/2$, and the qualitative conclusion remains for other reasonable removal rates.
Examining the density profile of topologically nontrivial configuration we find prominent occupancy at the boundary approaching the value predicted by $\mathcal{N}^2$ Figure~\ref{Fig:Depletion}(b,c).
The edge states are observable when the initial filling is above half-filling.
The topologically trivial configuration, in contrast, exhibits diminishing edge density as shown in Figure~\ref{Fig:Depletion}(d,e).
The comparison shows edge states are revealed in noninteracting systems with atom depletion away from the edge.

\begin{figure}
\includegraphics[width = \columnwidth]{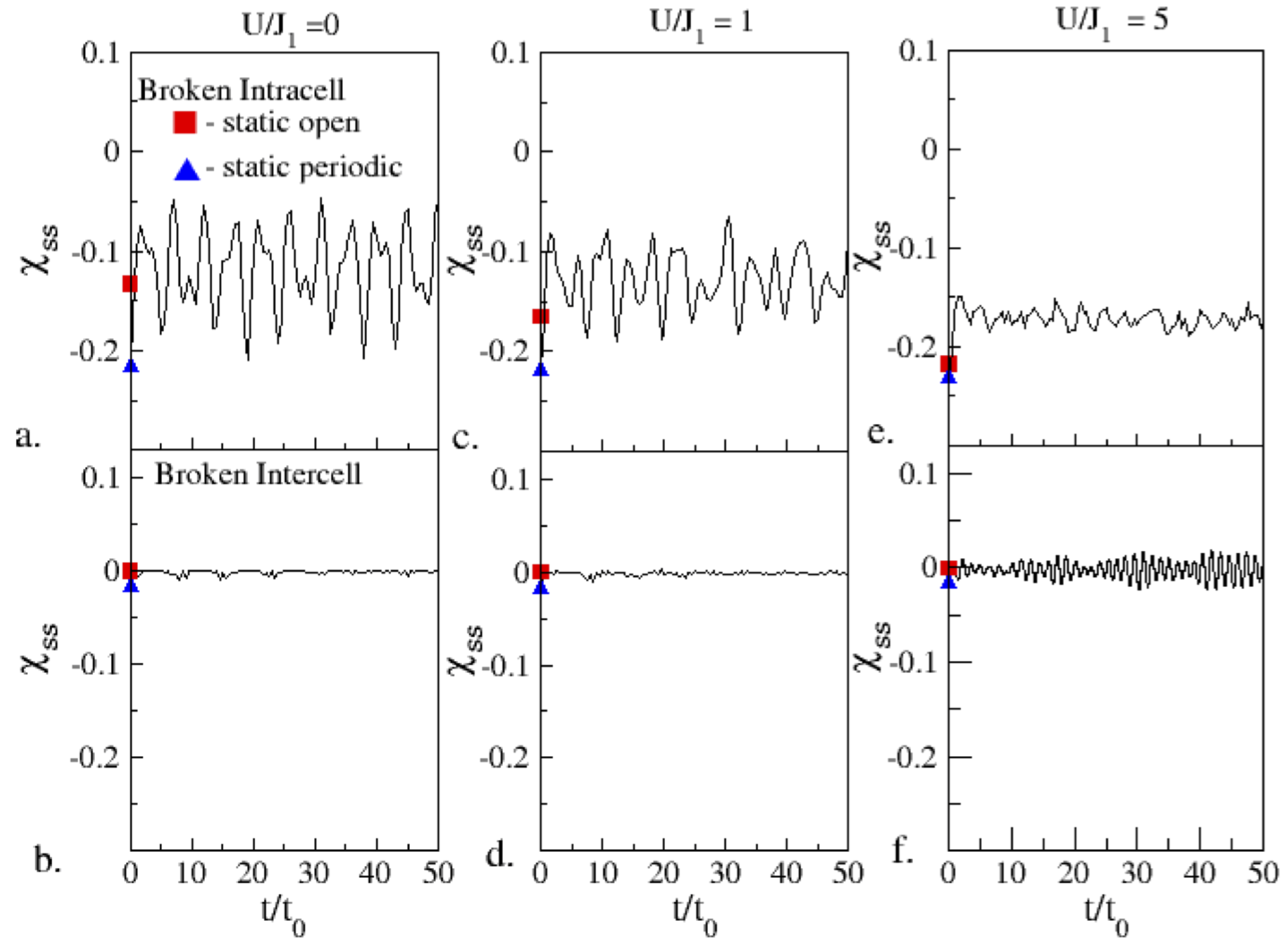}
\caption{Spin-spin correlations of fermions across a suddenly broken link of a dimerized, $L\!=\!12$ site lattice with $J_1/J_2\!=\!0.5$.
The topologically nontrivial configuration (top row) supports two edge modes in the noninteracting case with open boundary condition, and the cut is across a strong link $J_2$ for $U/J_1\!=\!0,1,5$ (a,c,e).
The correlations remain finite after the boundary transformation.
The topologically trivial configuration (bottom row) has a cut at a weak link $J_1$ for increasing interactions $U/J_1\!=\!0,1,5$ (b,d,f).
The correlations fluctuates around zero. The triangular (square) symbols show the time-independent correlation of a static configuration with periodic (open) boundary condition for the corresponding case.
In all cases the system is half-filled.}
\label{Fig:TDSSCorr}
\end{figure}

{\it Dynamical Signatures in Interacting Systems:}
Determining topological classification of systems with interactions is difficult. A periodic, dimerized lattice has nontrivial topological features in the presence of onsite interactions when particle-hole symmetry is respected~\cite{Hatsugai2006}. Therefore, distinguishing topological features of a dimerized lattice with interactions experimentally using cold atoms is desired. Rather than probing global topological number, specific short range correlation functions will indicate nontrivial topology and dimerization structures. Correlation function dynamics is observable as the boundary is changed from periodic to open. Dynamics of some correlations have been evaluated for both trivial and nontrivial topological configurations of the noninteracting SSH model and the Kitaev chain~\cite{He2016a}. Importantly, correlations of density or spin can be experimentally measured using phase-contrast imaging which can distinguish between populations of each component~\cite{PC-Image2006, Gerton2000, Liao2010}.

To model the contact interaction of atoms, we use the two-component Hubbard model with onsite repulsion described by $H_U = U \sum_i n_{i\uparrow}n_{i\downarrow}$.
Here $U$ is the coupling constant and $n_{i\uparrow}$ ($n_{i\downarrow}$) is the number operator for spin up (down) fermions.
The hopping Hamiltonian, Eq.~\eqref{eq:HSSH}, is generalized to two components as well. Here the two components $\sigma\!=\uparrow,\downarrow$ may be two different internal states of the same atomic species. The one-dimensional Fermi Hubbard model can be solved analytically using the Bethe-Ansatz revealing, in the thermodynamic limit, a half filled uniform lattice has a charge gap for any finite interaction~\cite{Lieb2003}.
As discussed previously, a dimerized lattice is a band insulator when $U=0$ at half-filling or full-filling, and it is a topological Mott insulating phase at half-filling when $U>0$~\cite{Manmana2012}. By choosing suitable correlation functions, it is possible to distinguish between the topologically nontrivial and trivial configurations in conjunction with band insulating and Mott insulating phases.

We use exact diagonalization (ED) to determine the ground state~\cite{Lin1993, Sharma2015} up to moderate system size and already observe prominent features in relatively small systems.
If larger systems are being considered, density matrix renormalization group (DMRG) method may be used~\cite{White1992, WhiteDMRG1993, DMRGRev}.
The results presented are from a lattice with $L\!=\!12$ sites, and the same behavior is already observable for systems as small as $L\!=\!8$ sites.
Dynamics of the system can be simulated efficiently by applying a Krylov-space approximation to the time evolution operator $U(t)\!=\!e^{-i\hat{H}dt/\hbar}$ to reduce the dimension of matrix operations~\cite{Manmana2005}.
The approximation is limited by the Hamiltonian's spectrum width, and we compared the results against the full, exact solution of small size systems to verify the accuracy.
The time evolution of correlation functions across a selected link can then be monitored.

To identify which correlation function can clearly reveal the dimerization, we investigate the equal time spin-spin correlation between the sites across the broken link (connecting sites $i$ and $i+1$):
\begin{equation}
\chi_{ss}(t) = \langle n_{i+1,\sigma}(t)n_{i,\sigma}(t) \rangle - \langle n_{i+1,\sigma}(t) \rangle \langle n_{i,\sigma}(t) \rangle.
\end{equation}
It probes the correlations between the same spin component across the selected link. If measured across a strong (weak) link with $J_2$ ($J_1$) and $J_1\!<\!J_2$, the correlation is intra-cell (inter-cell). Figure~\ref{Fig:TDSSCorr} shows the correlation in static configurations with open or periodic boundary condition (marked by the symbols). For the intra-dimer correlation, $\chi_{ss}(t)$ remains finite for both the noninteracting ($U/J_1\!=\!0$) and interacting ($U/J_1\!>\!0$) cases. In contrast, the correlation decays to zero for the topologically trivial case. The finite spin-spin correlation results from the spin singlet on the link $(|\uparrow\downarrow\rangle-|\downarrow\uparrow\rangle)$ preferred inside each dimer, but inter-dimer spin correlations should be weak because long-range spin order is forbidden in 1D~\cite{Mermin66} and there is no spin gap for a uniform chain in the exact solution~\cite{Lieb2003}. Spin correlation within a dimer is dominated by the singlet behavior and prefers anti-parallel spins, while the spin correlation between adjacent dimers is practically random. Meaning, the spin-spin correlation function can distinguish whether an inter-cell or intra-cell link has been broken, working for both noninteracting and interacting cases.

The dynamics of correlation functions can be further explored by setting the lattice configuration initially half-filled and periodic, then changing the boundary condition by suddenly cutting and measuring correlation across a link. In isolated system like cold-atoms trapped in optical potentials, there is no external energy dissipation. As shown in Figure~\ref{Fig:TDSSCorr}, the correlations oscillate around their average values, but the contrast between the finite correlation of the intra-cell case and the vanishing correlation of the inter-cell case persists in the time-evolution. The correlation can sometimes overshoot the equilibrium values in its time-evolution. Morevoer, the intra-cell correlation does not decay to zero and exhibits memory effects of its initial state. We mention that previous work has investigated long-range correlations in dimerized lattices by analyzing entanglement entropy~\cite{Jhu2016}, and the edges exhibit quantized Renyi entanglement entropy from an effective coupling of sublattices with finite size~\cite{Wang2015}.

The spin-spin correlations, however, do not distinguish between the (noninteracting) band and (interacting) Mott insulating phases of the dimerized lattice, and we need another correlation function to accomplish this. At half-filling, the Mott insulating phase suppresses onsite density fluctuations but the band insulating phase, formed by a pile of delocalized wavefunctions, does not. One can thus differentiate the two phases at half-filling by checking the onsite density fluctuation $\langle n_{i,\sigma}^2\rangle - \langle n_{i,\sigma}\rangle^2$, which vanishes in the Mott-insulator phase \cite{Deng15}.

{\it Experimental Implementation:} 
These protocols for detecting edge states in dimer ring lattice potentials can be implemented by combining techniques that have already been experimentally demonstrated. Bose condensates in ring potentials have been created in numerous experiments, including some with one or more controllable ``barrier'' potentials.~\cite{WrightPRL2013, Ryu2013}. Structured light beams suitable for creating one-dimensional ring lattices can be created using phase and amplitude modulation techniques~\cite{Franke-ArnoldOpticalOE07, Amico2014, Lee2014, KumarDopplerRing}, or using time-averaged ``painted'' potentials~\cite{PaintPot2009, MullerPhD2011}. High-quality, stable ring lattices are probably best created using azimuthal phase imprinting techniques. These techniques are limited to producing ring lattices with even numbers of sites. If ring lattices with odd numbers of sites are to be explored as well, it may be necessary to use pure amplitude modulation techniques in a system with higher optical resolution.

Even with current ``Fermi gas microscopes''~\cite{Bakr2009, MullerPhD2011, Greiner2015}, it is challenging to project a ring lattice potential where tunneling energies are sufficiently large. Using a low-mass atom such as $^6$Li may be essential for success. With azimuthal phase imprinting on a (780 nm) ``ring'' beam, an optical system with numerical aperture 0.6 can project a 32-site ring with lattice sites 1.4 $\mu$m apart. Only a few milliwatts of optical power would be needed for a lattice depth of 5$E_r$, which for $^6$Li would result in the tunneling energy $J\approx$ 1 kHz. Here $E_r$ is the recoil energy. (The tunneling rate for $^{40}$K would be 6.6 times smaller.) With micron-scale confinement in all (azimuthal, radial, axial) directions and the ability to tune the interaction energy via Feshbach resonance, on-site interaction strengths of 0-5$J$ are obtainable.

These protocols assume single-site detection capability, which has been demonstrated with both $^6$Li  and $^{40}$K~\cite{Cheuk2015}. Achieving low heating rates in a ring lattice with MHz scale trap frequencies, required for Raman sideband cooling, may be more challenging than creating a suitable potential for the main portion of the protocol. However, laser power requirements will be somewhat relaxed compared to 2D lattice experiments because of the much smaller trapping volume. We also note that site-resolved spin-sensitive (Faraday) imaging has been recently been reported (with Bosonic $^{140}$Yb)~\cite{Yamamoto2016}, and this technique might be adapted to resolve spin correlations in the proposed ring lattice.

{\it Discussion: } While solid state systems can reveal the influence of topological edge modes on transport, cold atoms can characterize topological properties directly associated with the wavefunction~\cite{Chien15}. Since cold-atoms need to be confined in experiments, the confining potential distorts the density profile and hinders direct observation of topological edge modes. We propose a geometry suitable for generating edge modes in 1D by cutting open a ring lattice, and two dynamical probes for revealing the edge modes by exploiting their localized nature: One utilizes single-particle injection in real space and the other depletes the particles away from the boundary.

Identifying topological properties of the interacting SSH model received theoretical progress for determining global behavior, but experimental verification can be challenging. Evaluating the time-evolution of selected correlation functions provides a local probe of dimerization signatures and distinguishes interacting and noninteracting systems in different configurations.
The setup and probes are general and should be useful for experimental verification of bulk-boundary correspondence and explore dynamics in other topological systems.

\bibliographystyle{apsrev4-1}
%

\end{document}